\documentclass[a4paper,aps,pre,twocolumn,groupedaddress,showkeys,showpacs,floats,floatats,floatfix,dvipdfm]{revtex4-1}
\usepackage[utf8]{inputenc}
\usepackage[english]{babel}
\usepackage{graphicx}
\usepackage{dcolumn}
\usepackage{bm}
\usepackage{natbib}
\usepackage{latexsym}
\usepackage{mathrsfs}
\usepackage{amssymb}
\usepackage{amsmath}
\usepackage{amscd}
\usepackage{color}
\usepackage{pstricks,pst-node,pst-text,pst-3d}
\usepackage{verbatim}
\usepackage[T1]{fontenc}
\newcommand{\RC}{${\cal RC}$}
\begin{document}
\title{Generic Structures in Parameter Space and Ratchet Transport}
\author{A.~Celestino$^1$, C.~Manchein$^{1,2}$, H.A.~Albuquerque$^1$ and M.W.~Beims$^{2}$}
\affiliation{$^1$Departamento de F\'\i sica, Universidade do Estado de
  Santa Catarina, 89219-710 Joinville, SC, Brazil}
\affiliation{$^2$Departamento de F\'\i sica, Universidade Federal do
    Paran\'a, 81531-980 Curitiba, PR, Brazil}
\date{\today}
%
\begin{abstract}
This work reports the existence of Isoperiodic Stable
Ratchet Transport Structures in the parameter spaces
dissipation {\it versus} spatial asymmetry and {\it versus} 
phase of a ratchet model. Such structures were found
[Phys.~Rev.~Lett.~{\bf 106} 234101 (2011)] in the
parameter space dissipation {\it versus} amplitude of the
ratchet potential and they appear to have generic shapes and
to align themselves along preferred directions in the parameter
space. Since the ratchet current is usually larger inside
these structures, this allows us to make general statements
about the relevant parameters {\it combination} to
obtain an efficient ratchet current. Results of the present
work give further evidences of the suggested generic
properties of the isoperiodic stable structures in the
context of ratchet transport.
\end{abstract}
%
\maketitle

Ratchet models are prominent candidates to describe the
transport phenomenum in nature in the absence of external bias. The
ratchet effect is a rectification of an external net-zero force to
obtain a directional motion of particles in spatially periodic media.
Spatiotemporal symmetries must be broken \cite{artuso07,wang07,flach00}
in the system
in order to obtain the {\it Ratchet Current} (\RC). In recent years
the literature related to the ratchet transport has increased enormously
as a consequence of promising applications. The \RC\, was observed
theoretically and experimentally in a variety of areas:
in Brownian \cite{polson10,cheng07,astumian02,linke02,reimann,magnasco}
and molecular motors \cite{julicher97}, cold atoms in optical ratchets
\cite{rietmann11,hagman11,zelan11,gommers08,flach08,flachOL06,dima05},
spin ratchets \cite{costache10}, flow density \cite{chen10},
organic electronic ratchets \cite{roeling11}, Levy ratchets 
\cite{pavlyukevich10,flach02}, Leidenfrost motion on ratchets 
\cite{lagubeau11,Stout}, granular gas \cite{meer10}, micro and nanofluids 
transport \cite{mathwig11,guo11,lambert10,hanggiFLUID08,ratchet-micro},
Fermi acceleration \cite{cesarMPE09}, coupled to finite baths
\cite{janePD09,janePRE,jane2,jane1}, magnetic films \cite{perez08},
energy transport \cite{flach06,flach05,flachPRL02},
classical and quantum ratchets in general
\cite{carlo11,dittrich01,kohler02,kohler,grifoni97}, among others.

Physical quantities like viscosity, particles mass, dissipation
$\gamma$, noise intensity, amplitude of external forces $K_t$,
ratchet potential amplitude $K$ etc., are usually, depending on the
system, the parameters which control the dynamics and the \RC.
Normally these quantities are deeply interconnected so that small
parameters variations may totally alter the dynamics and the
efficiency of the \RC. Thus it is very desirable, if possible, to
make general statements about the parameters {\it combination} which
generate large \RC s. An imperative development in this direction
suggested \cite{alan11-1} that {\it Isoperiodic Stable Structures}
(ISSs) in the parameter space present generic features which should
be valid for almost all inertia ratchet models, 
independent of their
application in nature. The ISSs are Lyapunov stable islands with
dynamics globally structurally stable. One of the generic features
found is the ``pattern'', or the ``shape'' of the ISSs. An example
of such generic shape is the shrimp-shaped ISS [see
Figs.~\ref{zoom1}(b) and \ref{zoom2}(b)], which has already appeared
in the parameter space of generic dynamical systems and applications
\cite{jasonPRL93,grebogi93,jason94,jason95,murilo96,beims-gallas97,
beims01,bonattoR07,bonatto08,stegemann10,vitolo11}, and observed
recently in experiments with electronic circuits
\cite{stoop10,viana10}. A second apparently generic feature is that
ISSs appear along preferred direction in parameter space, serving as
a guide to follow the \RC\, which is usually larger inside the ISSs.
In this perspective, the main goal of the present work is to show
that the ISSs found \cite{alan11-1} in the parameter space
($K,\gamma$) and ($K_t,\chi$) (this for the Langevin equation),
present the same generic behavior in another parameter spaces,
namely ($a,\gamma$) and ($\phi,\gamma$), where $a$ is the ratchet
asymmetry and $\phi$ is the phase of the ratchet potential. 
This is of relevance not only to show the generic properties of the
ISSs but for experimental realizations, where for specific experiments 
it is easier to vary different parameters. We show
a direct connection between \RC s with a family of ISSs and chaotic
domains in the above mentioned parameter spaces. 

The paper is organized as follows: Section \ref{model}
presents the model which will be used and in Sec.~\ref{ana} analytical
results are derived for the stability boundaries in parameter space
for fixed points. Section \ref{gammaK} shows the largest Lyapunov 
exponents
(LEs) in the parameter space ($K,\gamma$), results which were not
presented in \cite{alan11-1}, proving that the ISSs 
are stable. Sections \ref{gammaa}
and \ref{gammaphi} present the \RC, LEs and periods of the orbits in
the parameter spaces ($a,\gamma$) and ($\phi,\gamma$), respectively.
Section \ref{conclusions} summarizes the main results and gives some
perspectives for the experimental observation of ISSs.

\section{The Model}
\label{model}
In order to show generic properties of the ${\cal RC}$ in
the parameter space, we use a map M which presents all
essential features regarding unbiased current \cite{casatiPRL07}
\begin{eqnarray}
  M:\left\{
\begin{array}{ll}
  p_{n+1} = \gamma p_n + K[\sin(x_n)+a\sin(2x_n+\phi)], \\
  x_{n+1} = x_n + p_{n+1},
\end{array}
\right.
\label{map}
\end{eqnarray}
where $p_n$ is the momentum variable conjugated to $x_n$, $n=1,2,\ldots,N$
represents the discrete time and $K$ is the nonlinearity parameter. The
dissipation parameter $\gamma$ reaches the overdamping limit for $\gamma=0$
and the conservative limit for $\gamma=1$.
The ratchet effect appears due to the spatial asymmetry, which occurs with
$a\ne0$ and $\phi\ne m\pi$ ($m=1,2,\ldots$), in addition to the time reversal
asymmetry for $\gamma\ne1$. The ${\cal RC}$ of the above model was
analyzed in the dissipation range  $0\le\gamma<1$, for fixed $K=6.5$
\cite{casatiPRL07} and in the parameter space $0\le\gamma < 1$ and 
$0\le K\le 14$ \cite{alan11-1}. It was shown that close to the limit 
$\gamma=1$ the ${\cal RC}$ arises due to the mixture of chaotic motion 
with tiny island (accelerator modes) from the conservative case, while for 
smaller values of $\gamma$, chaotic and stable periodic motion (ISSs) 
generate the current.

The  ${\cal RC}$ is obtained numerically from
\begin{equation}
{\mathcal RC}=\frac{1}{M}
\sum_{j=1}^M\left[\frac{1}{N}\sum_{n=1}^N p_n^{(j)}\right],
\end{equation}
where $M$ is the number of initial conditions. At next we determine the
${\cal RC}$ for different parameter spaces always using a grid of
$600\times600$ points, $10^5$ initial conditions with $<p_0>=<x_0>=0$
inside the unit cell ($-2\pi,2\pi$) and $N=10^4$ iterations.

\section{Analytical boundaries for fixed points ($\phi=\pi/2$)}
\label{ana}
\begin{figure}[htb]
  \centering
  \includegraphics*[width=0.95\columnwidth]{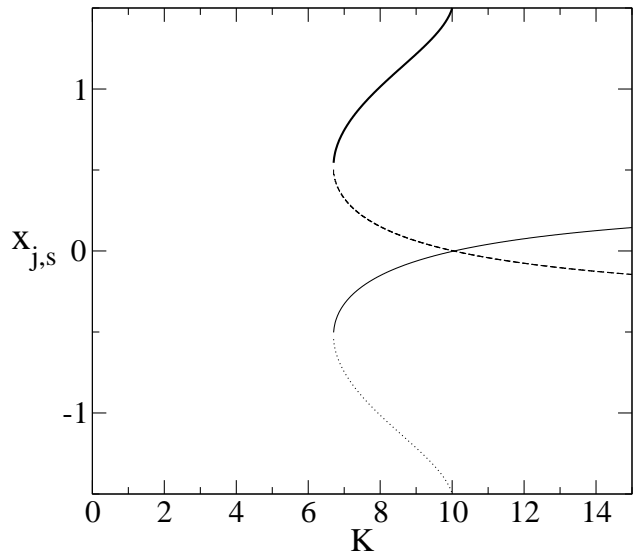}
  \caption{Fixed points $x_1$ (dashed line), $x_2$ (thin continuous), $x_3$ 
(thick continuous), $x_4$ (dot) as a function of $K$ for 
$a=0.5,\phi=\pi/2, \gamma=0.2$ and $L=1$ .}
  \label{x}
\end{figure}

Analytical boundaries for the  \RC\, in the parameter space can be
determined for fixed points. They are obtained from the analytical
expression for the eigenvalues of the Jacobian of the map (\ref{map})
after one iteration. Using $\phi=\pi/2$  the fixed points from
(\ref{map}) can be calculated from
\begin{eqnarray}
& & p^{(1)} - 2\pi L =0,\cr
& & \\
 &  &2\pi  L (\gamma-1) + K\left[\sin{(x^{(1)})}+ a\cos{(2x^{(1)})}\right]=0,
\nonumber
\end{eqnarray}
where $L$ is an integer or rational number. The orbital solutions are

\begin{eqnarray}
x^{(q=1)}_j & = &\arctan{(\alpha^{(-)},\pm \beta^{(+)})}, \quad (j=1,2) \cr
& & \label{xo}\nonumber \\
x^{(q=1)}_s & = & \arctan{(\alpha^{(+)},\pm \beta^{(-)})},\quad (s=3,4)
\nonumber
\end{eqnarray}
where

\begin{eqnarray}
\alpha^{(\mp)} & = & K\mp\sqrt{f},
     \qquad f= K \left[8Ka^2+K+16aL\pi(\gamma-1)\right],\cr
& & \cr
\beta^{(\pm)}  & = & \sqrt{8aL\pi(1-\gamma) + 4Ka^2 - K \pm\sqrt{f}}.
\label{xs}
\nonumber
\end{eqnarray}
The orbital points $x_{j,s}$ are plotted in Fig.~\ref{x} 
as a function of $K$, for $a=0.5,\phi=\pi/2, \gamma=0.2$ and $L=1$. 
Dashed and dot-dashed lines are respectively the unstable points 
$x_1,x_4$, while thin and thick continuous lines are respectively the 
stable points  $x_2, x_3$. We see that at $K=6.0$ all four fixed points
are born while for $K=10$ the stable and unstable points $x_3,x_4$ 
collide and vanish (observe that is mod $2\pi$). The points $x_1,x_2$ 
remain for higher values of $K$. Increasing the values of $L$, points 
$x_1,x_2,x_3,x_4$ move to higher values of $K$, but the bifurcation 
structure remains the same.

Substituting the solutions (\ref{xo}) in the Jacobian of the map (\ref{map}) 
after one iteration, we obtain analytical expressions for the eigenvalues 
$\lambda(K,\gamma,L)$.
These expressions are very large and are not written explicitly here.

\subsection{Born of period-$1$, any $L$.}
When the eigenvalues satisfy the relation $\lambda(K,\gamma,L)=+1$,
fixed points (or period-$1$ orbits) are born in the phase space. From
this relation it is possible to determine $\gamma(K,L)$,
which defines the boundaries in the parameter
space for which fixed points are born. The solutions for the
equation $\lambda(K,\gamma,L)-1=0$, for all fixed points $x^{(q=1)}_j$,
are
\begin{eqnarray}
\gamma_{\pm 1}^{(1)}(L\ne0,K,a) & = &1+\frac{K}{2\pi L}(a\pm1),\\
& & \label{gi}\cr
\gamma_{2}^{(1)}(L\ne0,K,a) & = & 1 -\frac{K}{2\pi L}\left(a+\frac{1}{8a}\right).
\label{gii}
\end{eqnarray}
For $L=0$ we found just the solution $a=1$, independent of $\gamma$
and $K$.

\subsection{Bifurcation-$1\to 2, \, L=0$.}
When the eigenvalues satisfy $\lambda(K\gamma,L)=-1$, a double period
bifurcation $1\to 2$ occurs in phase space. For $L\ne0$ we were not
able to find analytical expressions for $\gamma(K,L)$. However, for
$L=0$ the analytical solution are given by
\begin{equation}
\gamma^{(1\to2)}(K,a) = -1 +
\frac{K}{8a}\left(\xi\sqrt{\xi^2-2\xi-3} \right),
\label{gamma12}
\end{equation}
where $\xi=\sqrt{8a^2+1}$.

The curves $\gamma_{\pm 1}^{(1)}, \gamma_{2}^{(1)}$ and $\gamma^{(1\to2)}$
define exactly sharp period-$1$ boundaries in the parameter space.
They generalize the curves shown in \cite{alan11-1} to any $K$ and $a$
values and will be presented later together with numerical results. We
were not able to find analytical solutions for the boundaries as a 
function of arbitrary $\phi$.

\section{Parameter space (${\bf K,\gamma}$) }
\label{gammaK}
\begin{figure}[htb]
  \centering
  \includegraphics*[width=0.95\columnwidth]{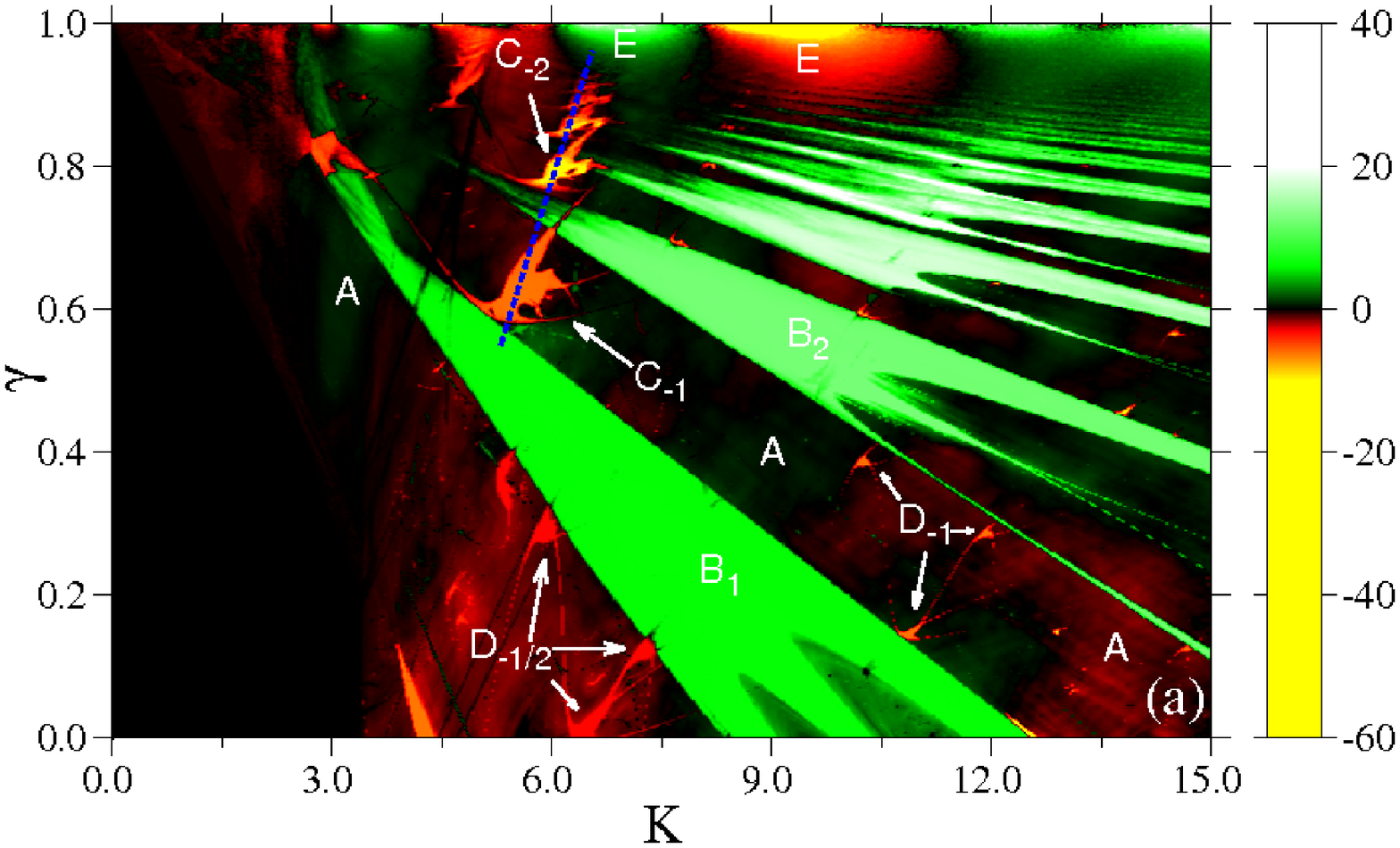}
  \includegraphics*[width=0.95\columnwidth]{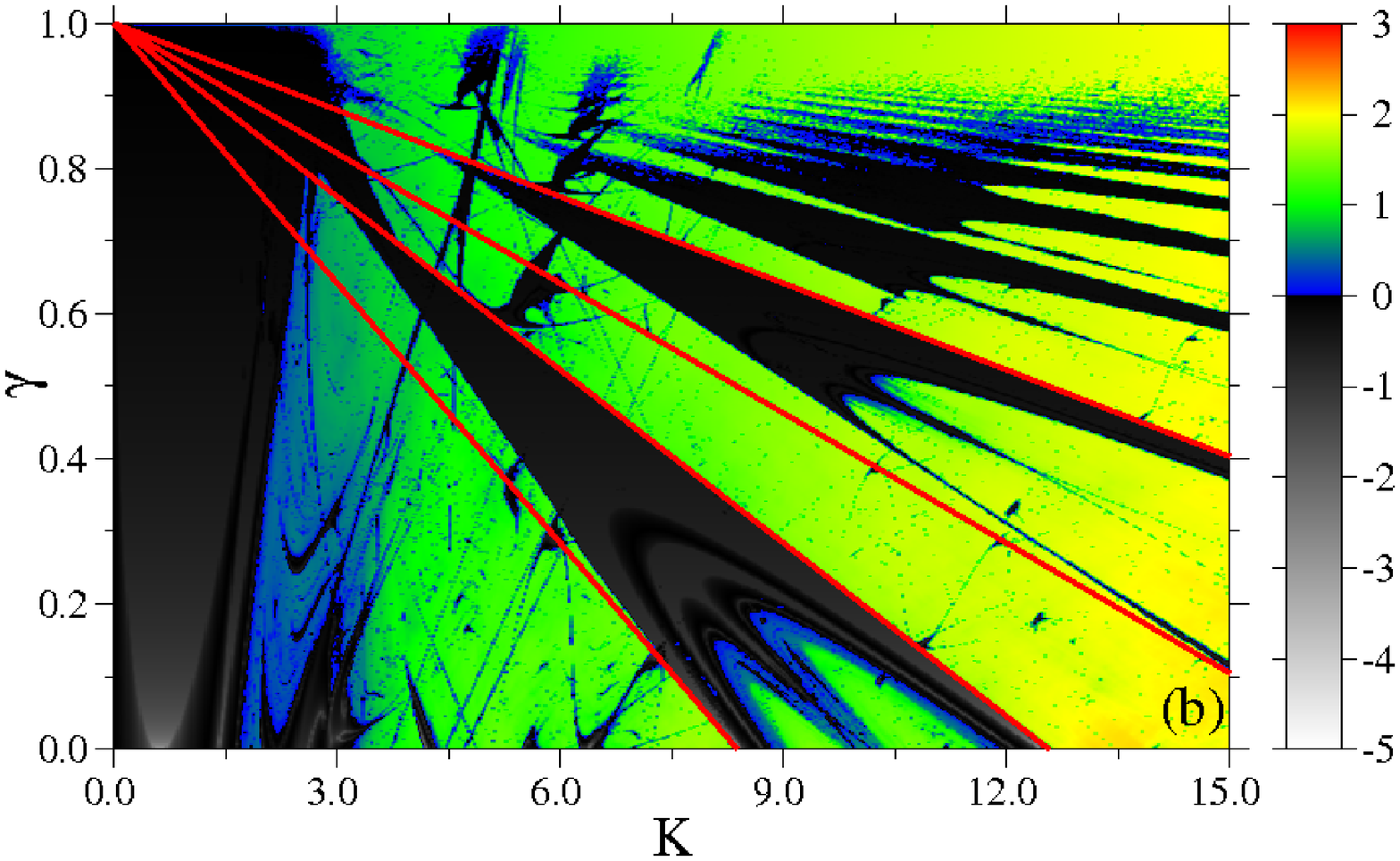}
  \caption{(Color online) (a) The ${\cal RC}$ and (b) largest Lyapunov 
  exponent plotted in the  parameter space ($K,\gamma$).}
  \label{PSS1}
\end{figure}

The purpose of this section is to show that the ISSs presented
in \cite{alan11-1} are {\it stable}. To do this we present the
LEs in parameter space. For comparison, we start showing in
Fig.~\ref{PSS1}(a) the ${\cal RC}$ plotted
in colors for the parameter space ($K,\gamma$) with $\phi=\pi/2$
and $a=1/2$. This is the parameter space analyzed in \cite{alan11-1}
but for a larger $K$ interval. While black colors represent
close to zero currents, red to yellow colors are related
to increasing negative currents. Green to white colors
are related to increasing positive currents (see color bar). An
unusual complex structure of colors is evident. Three distinct 
behaviors can be identified: ($i$) a large
``cloudy'' background, identified as $A$ in Fig.~\ref{PSS1}, mixed
with black, red and green colors, showing a mixture of zero, small negative
and positive currents, respectively; ($ii$) ISSs $B_L$ (cusp-like), $C_L$ and
$D_L$ (shrimp-like) embedded in the cloudy background regions;
($iii$) strong positive and negative currents (region $E$), with not well
defined borders which occur close to the conservative limit $\gamma=1$.
All ISSs are related to periodic \cite{alan11-1} and stable motion. The
stable behavior can be seen in
Fig.~\ref{PSS1}(b), which is the parameter space for the LE. Black to
white are related to negative LEs and blue, green, yellow to red related
to increasing LEs. All ISSs from Fig.~\ref{PSS1}(a), where the \RC\,
becomes usually larger inside, can be identified with the negative LEs
regions from Fig.~\ref{PSS1}(b). Thus the ISSs maintain their ``shape''
{\it independently} if we measure the \RC\, or the LEs in the parameter 
space. The same is observed for the periods of the orbits, shown in
Fig.~2 from \cite{alan11-1}. All analytical boundaries for the $B_L$
ISSs are obtained directly from $\gamma_{-1}^{(1)}(L\ne0,K,a)$
and $\gamma_{2}^{(1)}(L\ne0,K,a)$ given in Sec.~\ref{ana}. Four of them 
are plotted as red straight lines in Fig.~\ref{PSS1}(b).

In the cloudy background region $A$ from Fig.~\ref{PSS1}(a), smaller \RC s
are observed and can be directly identified with the regions of
positive LEs from Fig.~\ref{PSS1}(b). These \RC s are the consequence of
the asymmetry of the chaotic attractor \cite{dima05}. Thus we clearly
see that the cloudy chaotic background is not efficient to generate the
${\cal RC}$ as the ISSs are, and that the magnitude of the
positive LEs do not change the values of the corresponding \RC s.
For region $E$ the \RC\, in Fig.~\ref{PSS1}(a) is shown to be enhanced, but
the structures does not have well defined borders. When compared to
Fig.~\ref{PSS1}(b) we observe that the region $E$ is chaotic. The origin
of the large \RC\, is due to accelerator modes which exist in the
conservative limit $\gamma=1$ and are responsible for the asymmetry
of chaotic attractors.

\section{Parameter space ($a,\gamma$)}
\label{gammaa}
\begin{figure}[htb]
  \centering
 \includegraphics*[width=0.95\columnwidth]{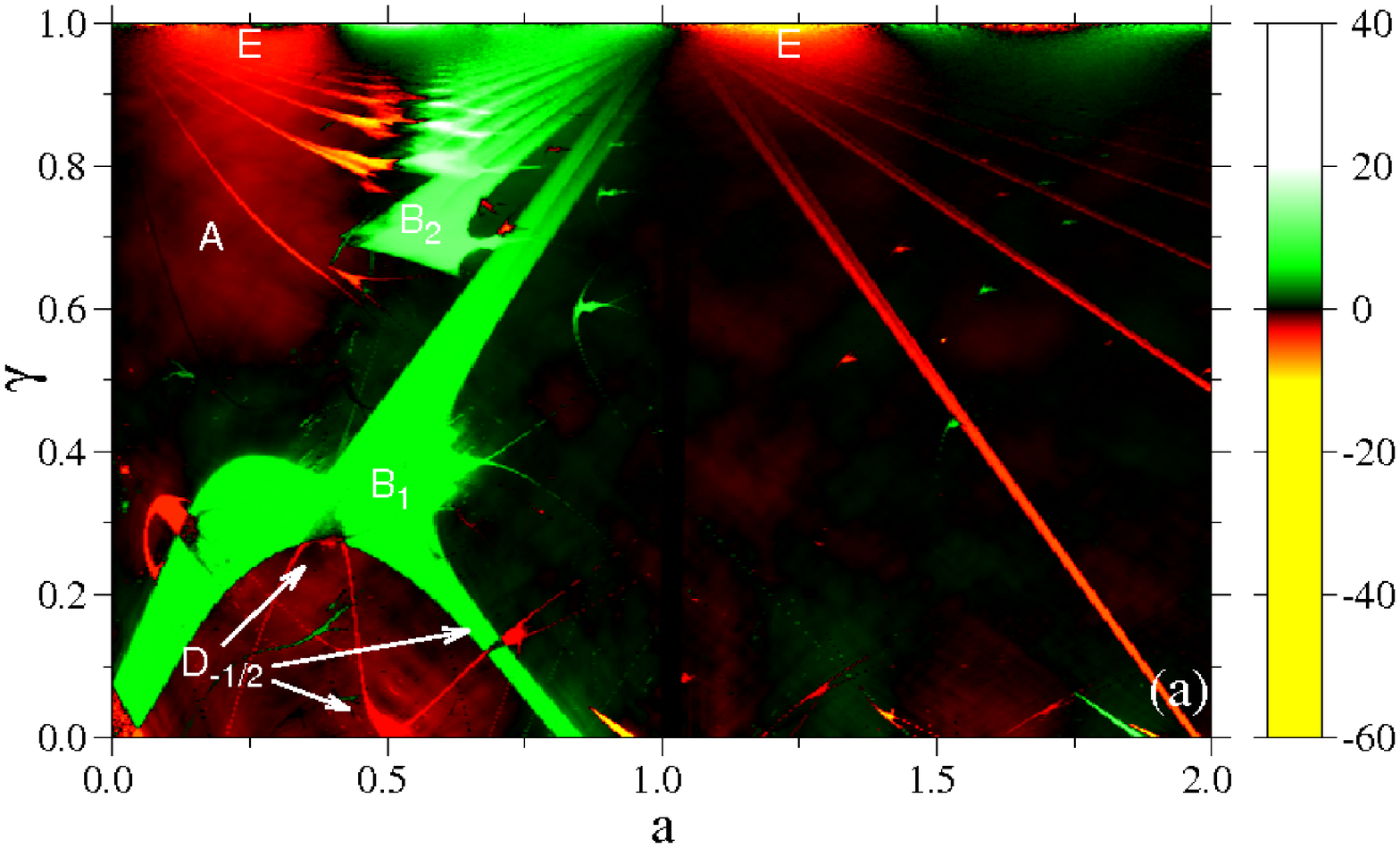}
 \includegraphics*[width=0.95\columnwidth]{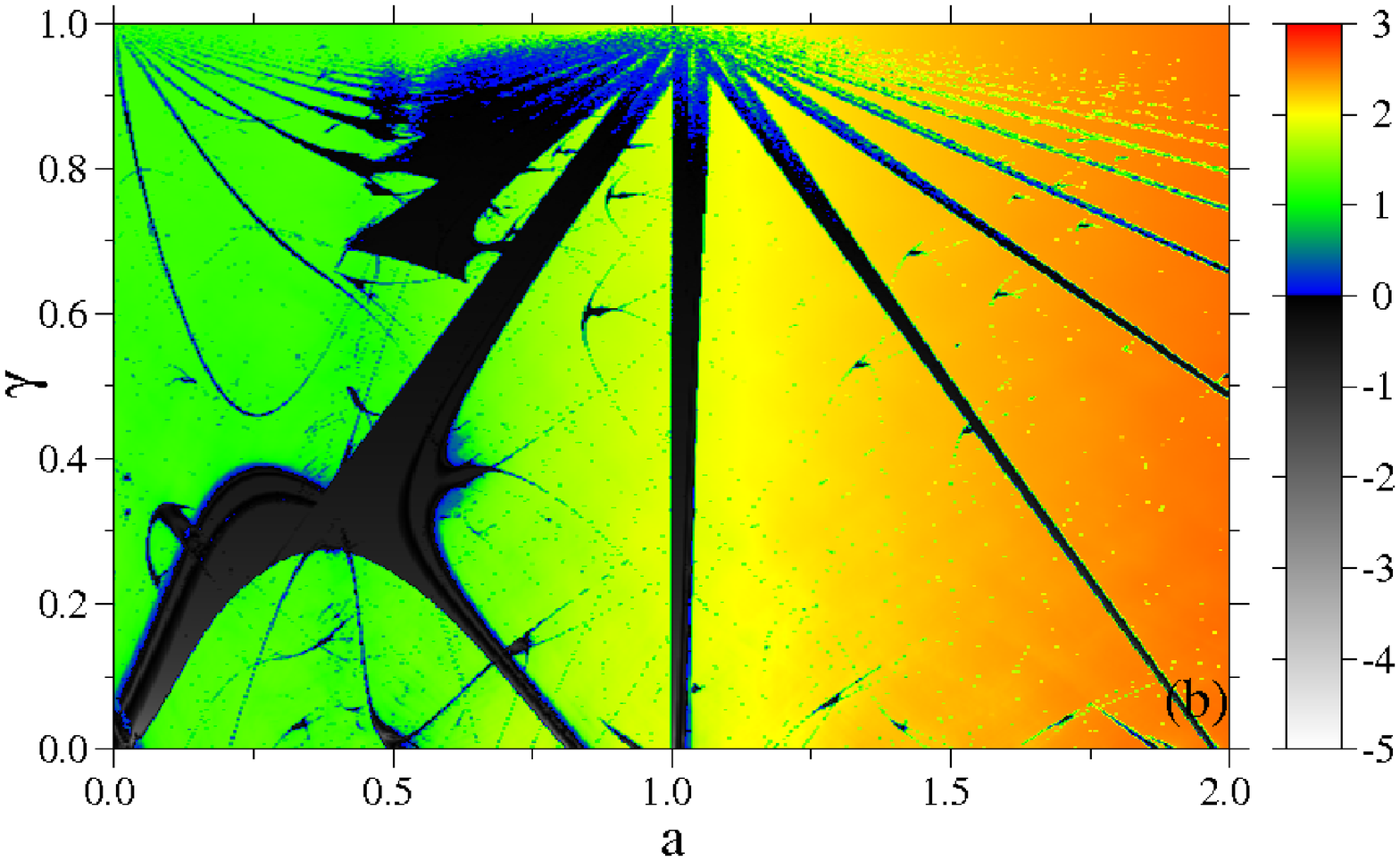}
 \includegraphics*[width=0.95\columnwidth]{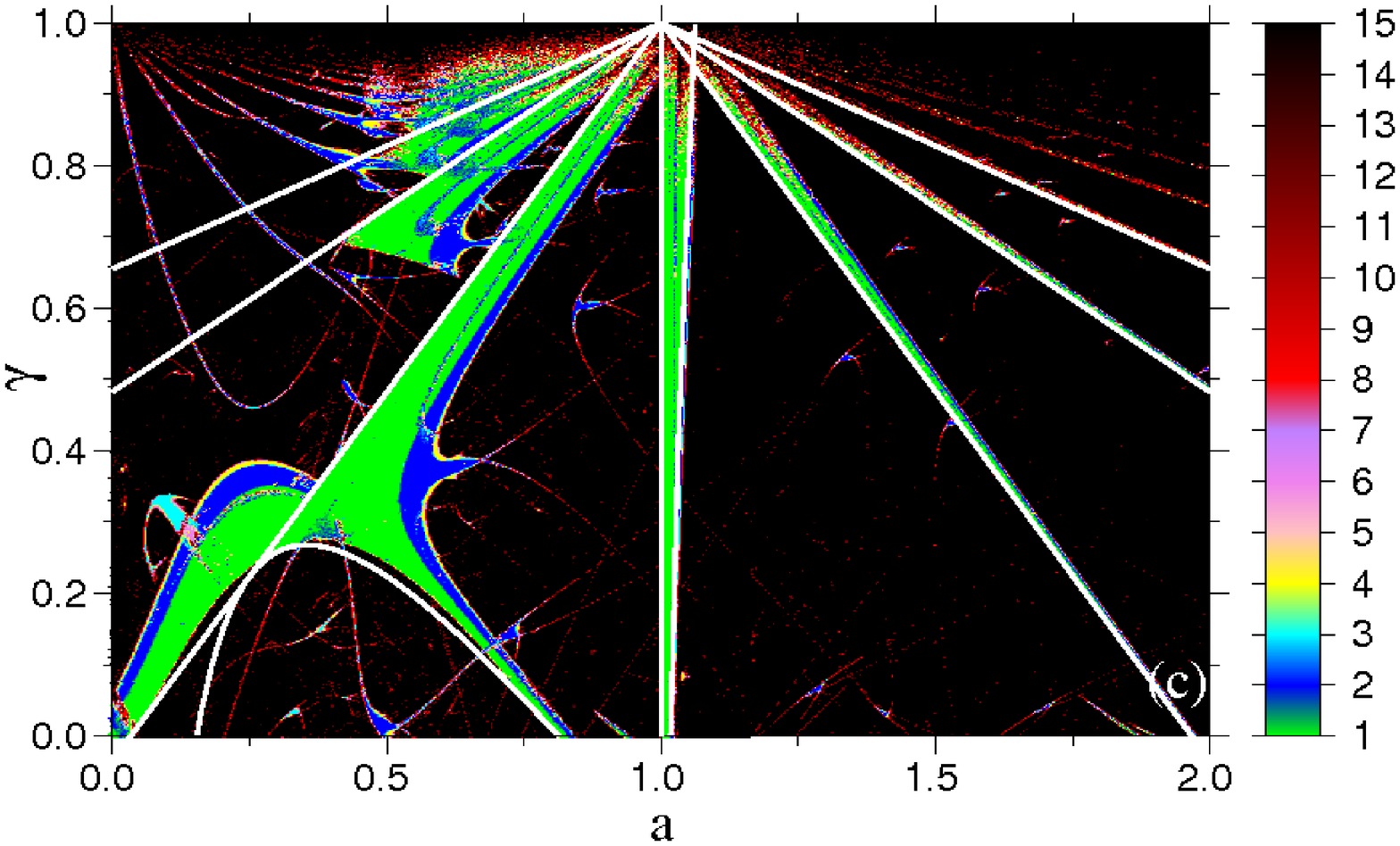}
 \caption{(Color online) (a) The ${\cal RC}$, (b) largest Lyapunov exponent
and (c) period $q$ of the orbits (see text) plotted in the  parameter
space ($a,\gamma$) for $K=6.5$ and $\phi=\pi/2$.}
  \label{a}
\end{figure}

The \RC\, in the parameter space ($a,\gamma$) is shown in
Figure \ref{a}(a) (see colorbar). Three main regions, as observed in
Fig.~\ref{PSS1}(a), are identified: the large cloudy background
region $A$, mixed with black, red and green colors,
showing a mixture of zero, small negative and positive currents,
respectively; ISSs with well defined borders
and embedded in the cloudy background region; and finally region $E$
with strong positive and negative currents with not well defined borders
which occur close to the conservative limit $\gamma=1$. Figure \ref{a}(b)
shows the corresponding parameter space ($a,\gamma$) for the LEs, where
black to white are related to negative LEs and blue, green, yellow to red
related to increasing LEs. Fig.~\ref{a}(c) shows the corresponding parameter
space for the period $q$ of the orbits. Periods $q$ are identified by green:
$q = 1$, blue: $q = 2$, cyan: $q = 3$, yellow: $q = 4$, pink: $q=6$,
red $q\ge 8$ and black for no period. In Fig.~\ref{a}(c) only one initial
condition is used ($x_0=0.5,p_0=0.3$), $N=10^6$ iterations and a grid of
$600\times600$ points. Comparing Figs.~\ref{a}(a),(b) and (c) a direct
connection between the ${\cal RC}$ with the chaotic and the regular
periodic behavior can be made. All ISSs appear with negative (black to
white) LEs and periodic motion while the cloudy background with small
values of the ${\cal RC}$s and region $E$, are related to the chaotic
motion and no periodic motion.

Larger green ISSs are observed for $a<1$, 
which apparently are responsible for the large positive \RC s. Thinner
red ISSs exist for $a>1$. Close to $a=1$ we also observe to have
a ISS with zero \RC. Inside all these ISSs a period doubling
bifurcation cascade
$1\times 2^n\, (n=1,2,\ldots)$ occurs, always starting from the left,
as can be observed in Fig.~\ref{a}(c). These left boundaries, where
fixed points are born, were obtained analytically in Sec.~\ref{ana} 
and some of them are plotted as white line in Fig.~\ref{a}(c). When 
using $L=1,2,3,\ldots$ in $\gamma_{-1}^{(1)}$ from Eq.~(\ref{gi}), we 
obtain {\it all} left boundaries for $a<1$, while using 
$L=-1,-2,-3,\ldots$ in $\gamma_{-1}^{(1)}$ we obtain {\it all} left
boundaries for $a>1$. For $L=0$ we also found the left boundary located
at $a=1$, which has current exactly zero inside. For this case with $L=0$
we were able to find the boundary of bifurcation $1\to2$, given by
Eq.~(\ref{gamma12}) and is plotted as a solid white line in Fig.~\ref{a}(c).
The \RC\, is exactly zero inside the ISSs with $L=0$ because points of
the periodic attractor are located exactly symmetrically around zero,
independent of the period. There occurs a {\it whole} symmetric period 
doubling bifurcation inside this ISS, for any dissipation value. 
Symmetric period doubling bifurcation means that orbital
points are symmetrically located around $p=0$, thus the \RC\, is zero.
ISSs with analytical boundaries clearly have
the $B_L$ cusp shape, where the inner part of the cusp is stretched out.
Going to the conservative limit $\gamma\to 1$ where $L\to\pm\infty$,
it is easy to show analytically, using results from Sec.~\ref{ana},
that the approximation rate $(B_{L+1}-B_L)/(B_L-B_{L-1})$
approaches $1$.

The other ISSs, namely $C_L$ and $D_L$ (shrimp-like), also appear
in the parameter space ($a,\gamma$). The shrimp-shaped ISSs are
immersed in the cloudy background region $A$. There are many of
such structures as can
be better seen in the LE parameter space from Fig.~\ref{a}(b). As
$\gamma$ decreases these ISSs follow preferred directions.
\begin{figure}[htb]
  \centering
  \includegraphics*[width=0.95\columnwidth]{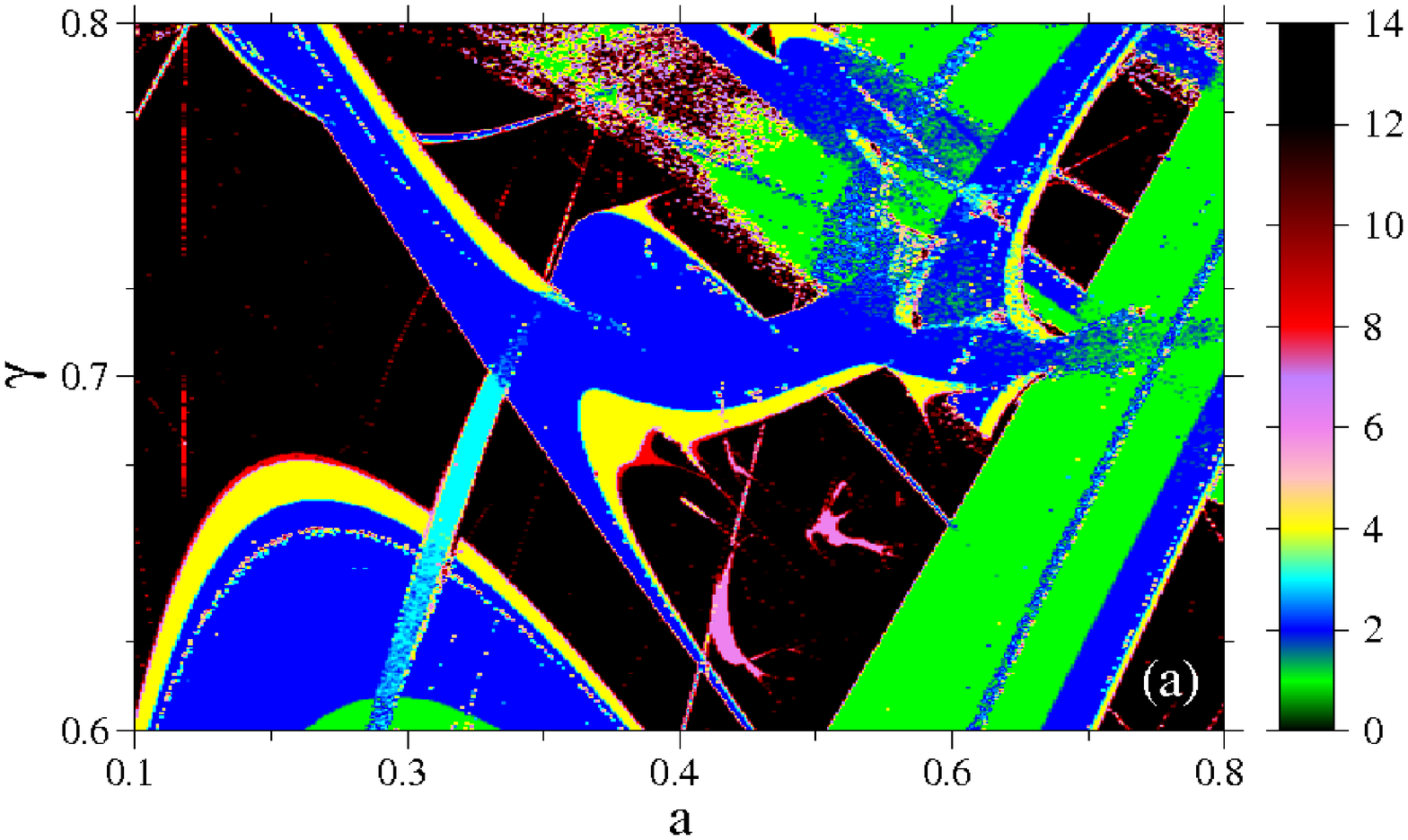}
  \includegraphics*[width=0.95\columnwidth]{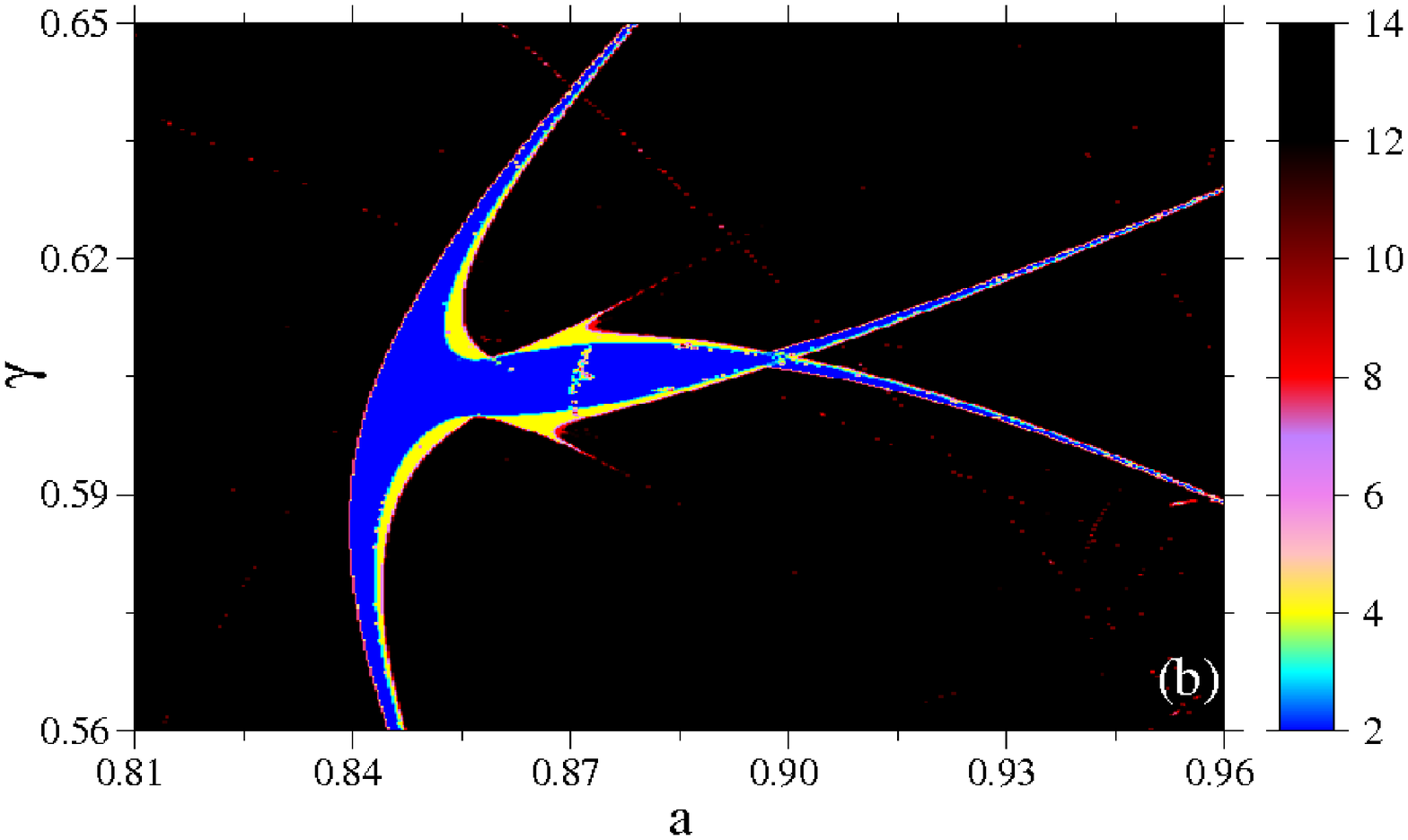}
  \caption{(Color online) Magnifications of Fig.~\ref{a}(c) showing
(a) $C_L$ and (b) $D_L$  (shrimp) ISSs.}
  \label{zoom1}
\end{figure}
The $C_L$ ISSs are harder to be observed, but can be recognized in the
magnification shown in Fig.~\ref{zoom1}(a). In this case we used $K=5.6,
\phi=\pi/2-0.3$, which is more appropriate to recognize the 
$C_L$ structures. In fact, we found out that the $C_L$ ISS is an overlap of 
the shrimp-like ISS and the inverted cusp \cite{jason06}. As $\gamma,K$ 
(and $\phi$) change they may start to go apart (close) and separate (join).
This is similar to what occurs for the shrimp-like structure, which
is an overlap of the cusp with the inverted cusp, which are generated 
by cubic maps \cite{lorenz08,cabral93}. On the other hand,
Fig.~\ref{zoom1}(b) shows a magnification of the typical shrimp-shaped 
ISS.

Also in the parameter space ($a,\gamma$) we observe that the magnitude
of the LE does not increase/decrease the \RC\, in region $A$. Compare
the LEs for $a<1$ with those for $a>1$, they are distinct but the
corresponding \RC s are essentially equal. Close to the conservative 
limit $\gamma=1$, we again observe a chaotic motion, as in the case 
of  Fig.~\ref{PSS1}(a), with larger \RC s, which are 
generated due to the mixture of the chaotic motion with the 
transporting islands from the conservative limit.

\section{Parameter space ($\phi,\gamma$)}
\label{gammaphi}
\begin{figure}[htb]
  \centering
 \includegraphics*[width=0.95\columnwidth]{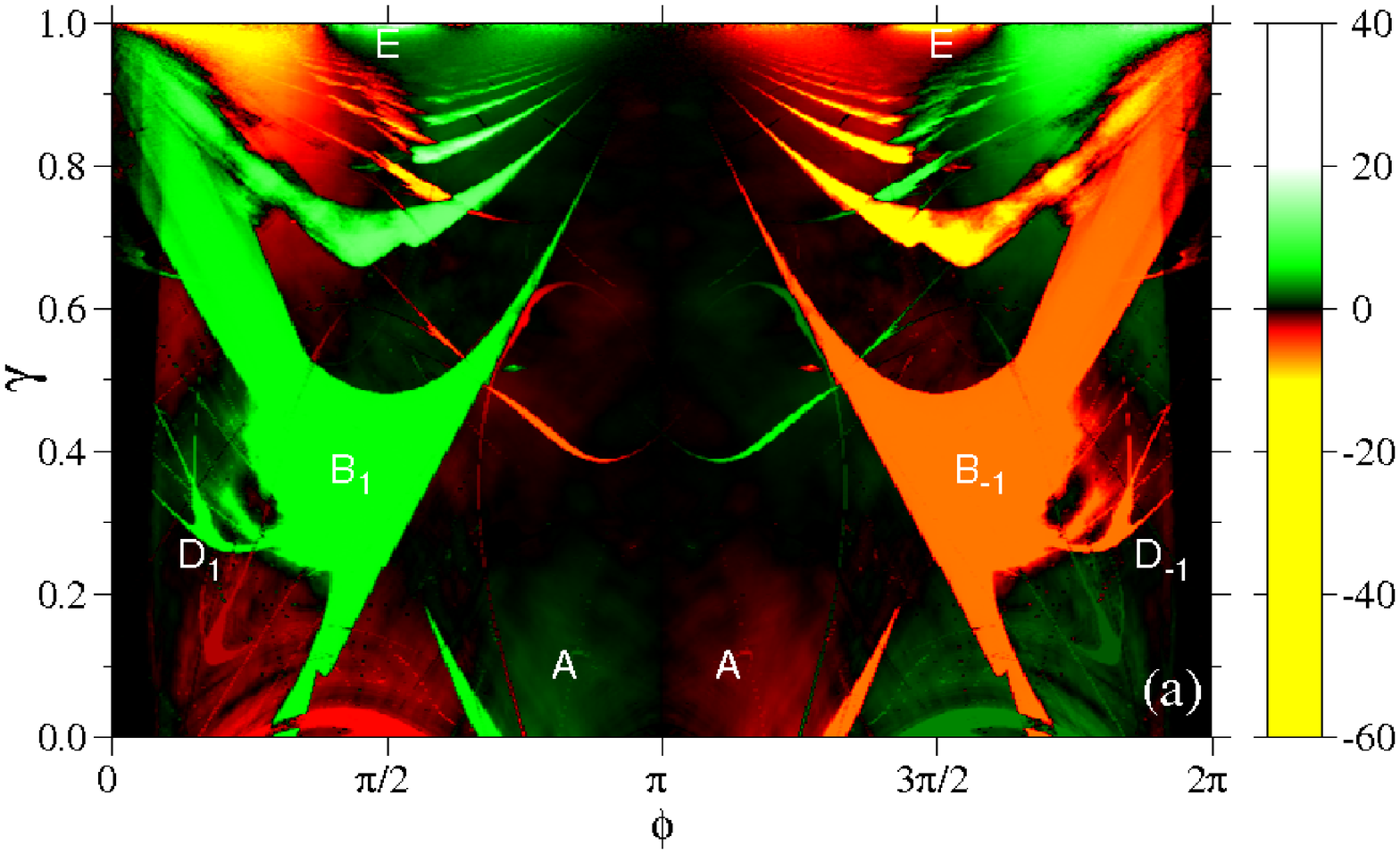}
 \includegraphics*[width=0.95\columnwidth]{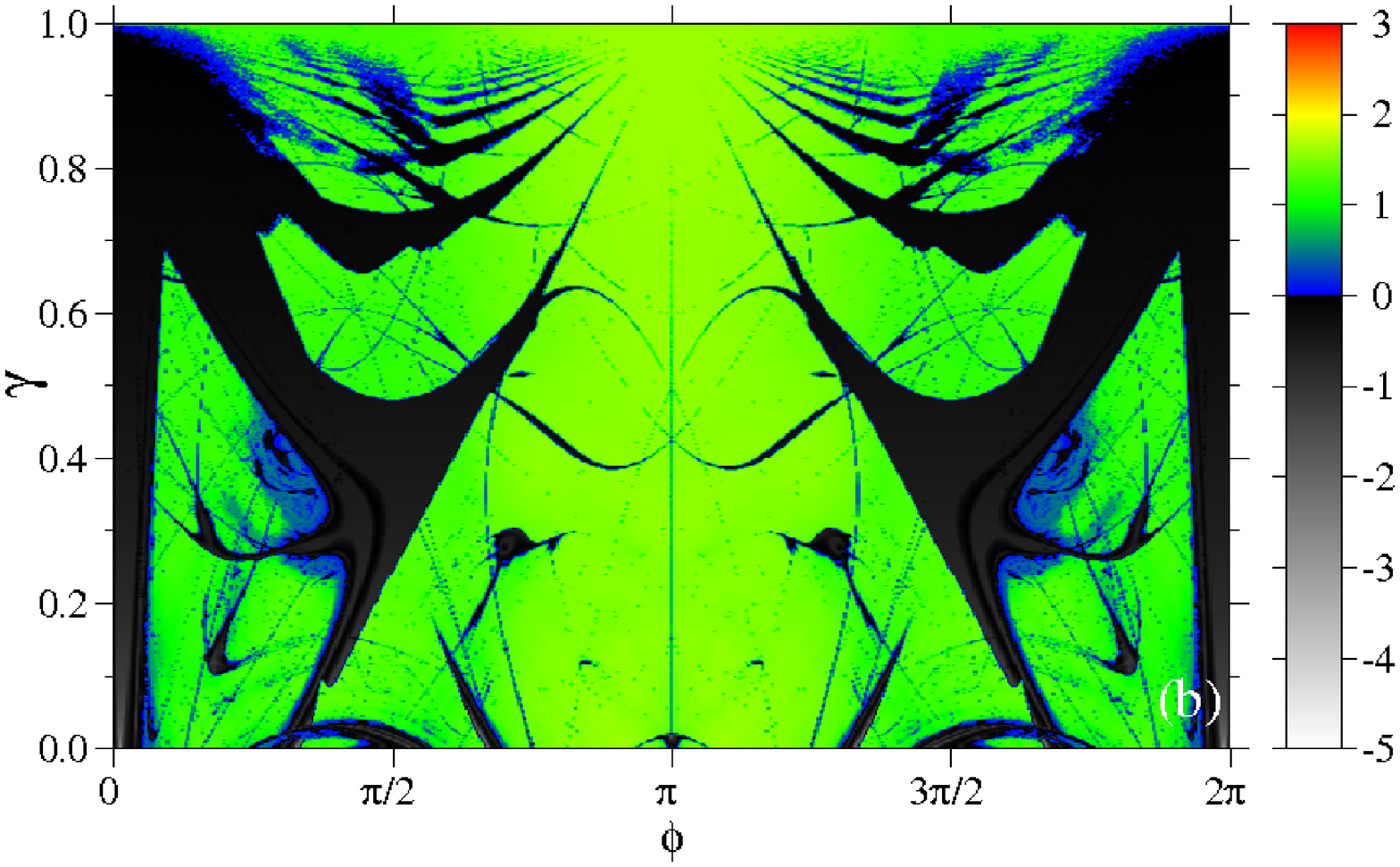}
 \includegraphics*[width=0.95\columnwidth]{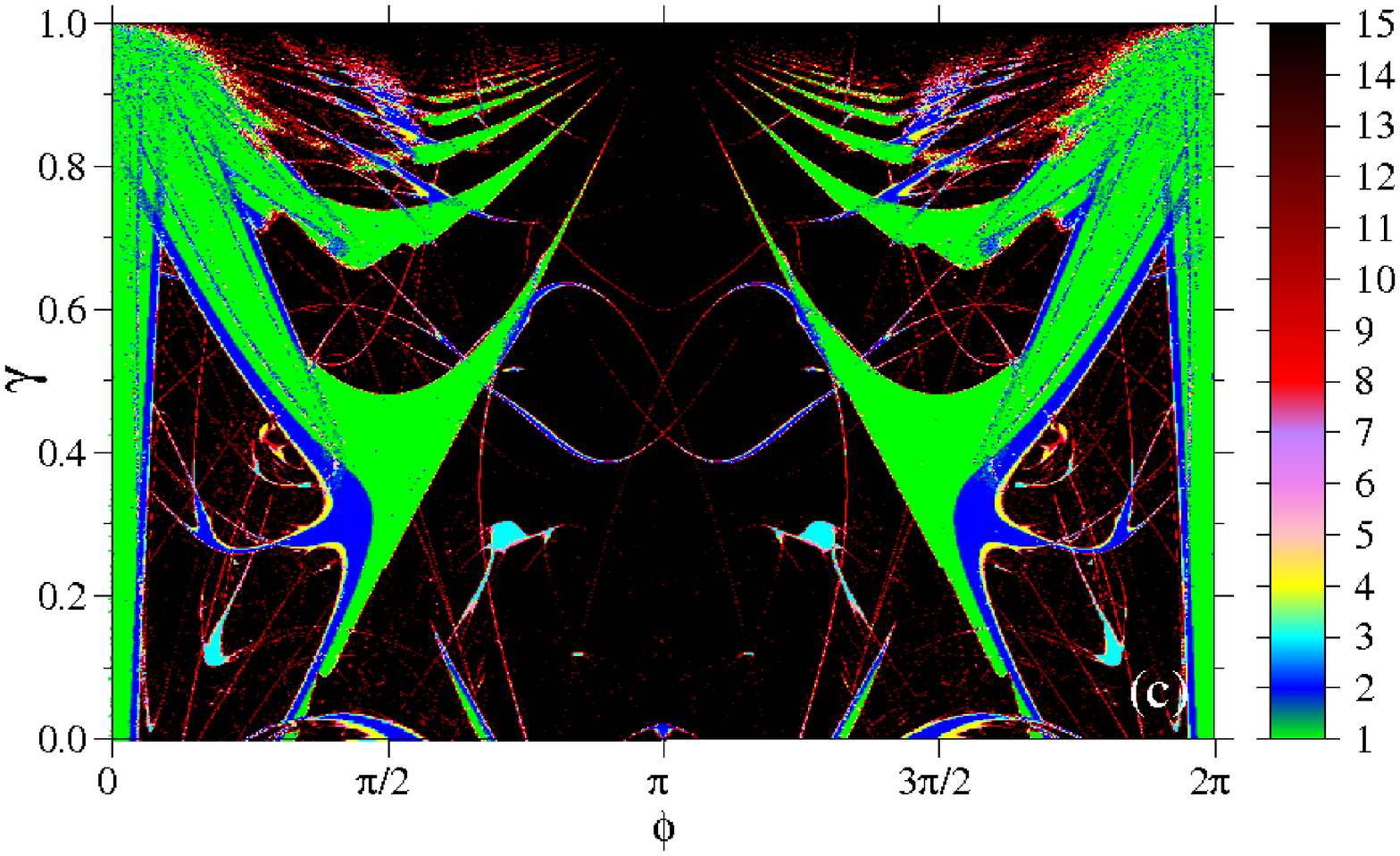}
 \caption{(Color online) (a) The ${\cal RC}$, (b) Lyapunov exponents
  and (c) period $q$ of the orbits (see text) plotted in the  parameter
  space ($\phi,\gamma$) for $K=6.5$ and $a=0.5$.}
  \label{phi}
\end{figure}

Figure \ref{phi}(a) shows the parameter space
($\phi,\gamma$) for the ${\cal RC}$. First observation is
that the ${\cal RC}$ is anti-symmetric under the transformation
$\phi\to 2\pi-\phi$, and thus current reversals occur when $\phi\to
2\pi-\phi$. A nice unusual structure is observed.
As for all other parameter spaces shown here, Fig.~\ref{phi}(a) displays
three main regions: the large cloudy background region $A$, mixed with
black, red and green colors,
showing a mixture of zero, small negative and positive currents,
respectively; the ISSs with well defined borders
and embedded in the cloudy background region; and finally region $E$
with strong positive and negative currents with not well defined borders
which occur close to the conservative limit $\gamma=1$.  Figure \ref{phi}(b)
shows the parameter space ($\phi,\gamma$) for the LE (see colorbar) while
Fig.~\ref{phi}(c) displays the corresponding periods of the orbits.
Comparing with Fig.~\ref{phi}(a) a direct connection of the ${\cal RC}$
with the regular periodic and chaotic behavior can be made. All ISSs appear
with negative (black to white) LEs (Fig.~\ref{phi}(b)) and periodic motion
in Fig.\ref{phi}(c), while the cloudy background with small
values of the ${\cal RC}$s and region $E$ are related to chaotic and
non-periodic motion. The magnitude of the LE does not increase/decrease the
\RC\, in region $A$. Close to the conservative limit $\gamma=1$ 
the large \RC s are again generated due to the mixture of the chaotic motion 
with the transporting islands from the conservative limit.

In this case it was not possible to determine analytically the boundaries
of the cusp ISSs. However, combining Figs.~\ref{PSS1},\ref{a} and \ref{phi}
it is possible to recognize that the large red (green) ISSs from 
Fig.~\ref{phi}(a), which look very similar to ``rips'', belong to the cusps. 
The first (from below) and larger cusp appears deformed and connected to 
shrimps [see white box
in Fig.~\ref{phi}(c)]. As $\gamma\to1$ these ISSs approach to each other.
\begin{figure}[htb]
  \centering
  \includegraphics*[width=0.95\columnwidth]{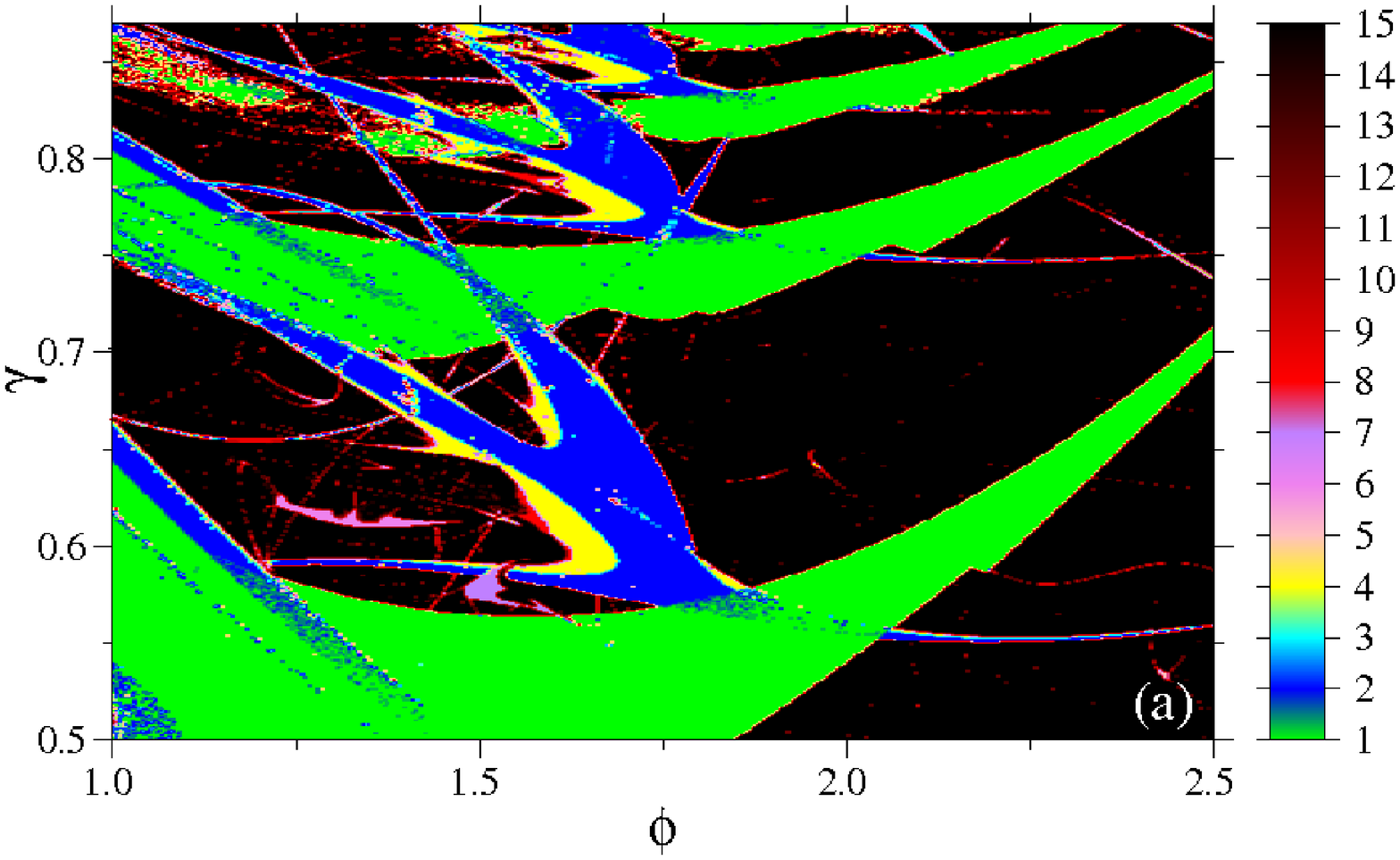}
  \includegraphics*[width=0.95\columnwidth]{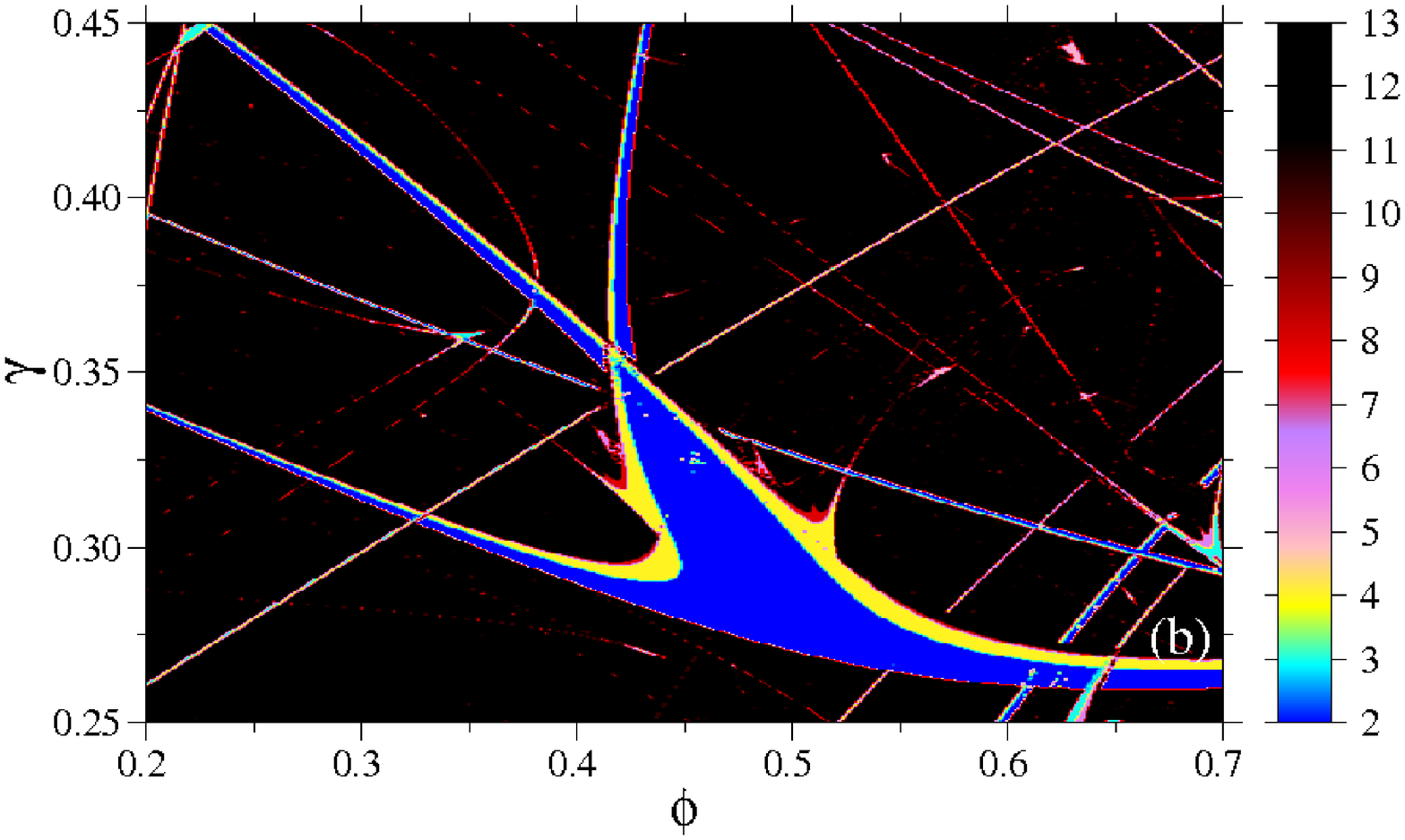}
  \caption{(Color online) Magnifications of Fig.~\ref{phi}(c) showing
(a) $C_L$ and (b) $D_L$ (shrimp) ISSs.}
  \label{zoom2}
\end{figure}
The $C_L$ and $D_L$ (shrimp-like) ISSs also appear in the parameter
space ($\phi,\gamma$) as can be recognized  in the magnifications
shown in Fig.~\ref{zoom2}. For better visualization of the $C_L$ ISS,
the magnification from Fig.~\ref{zoom2}(a) was calculated along the 
line $K = (\gamma + 0.994925)/0.2845$. This line was obtained from
Fig.~\ref{PSS1} by making a fit (see blue dotted line) along the 
preferred direction of the 
ISSs in that parameter space. On the other hand, Fig.~\ref{zoom2}(b) 
shows a typical $D_L$ shrimp ISS. Besides the above properties, the 
three plots show a remarkable combination of ISSs structures showing 
the rich and controlled dynamics which may be generated by tailoring 
the parameters.

\section{Conclusions}
\label{conclusions}
It was suggested recently \cite{alan11-1} that ISSs, inside which
the \RC\, is usually larger, should present generic features in
almost all parameter spaces of ratchet systems which display
dissipation $\gamma$ against a parameter related
to the ratchet property. In that work two parameter spaces
were analyzed: ($K,\gamma$) for the discrete model (\ref{map}),
where $K$ is the amplitude of the ratchet potential and ($K_t,\chi$) 
for the Langevin equation, where
$K_t$ is the amplitude of the external time dependent oscillating
force with zero mean. In the present work we show that the ISSs
appear also in the parameter spaces ($a,\gamma$) and ($\phi,\gamma$),
where $a$ and $\phi$ give, respectively, the spatial asymmetry and phase
of the ratchet potential (see Eq.~\ref{map}). We show a direct
connection between \RC s with a family of ISSs and chaotic domains
in the above mentioned parameter spaces. In all cases studied we observe 
that the ISSs usually follow three important properties: (1) their 
{\it pattern} (shrimps, cusps, etc..) remains the same, independent of 
the parameter space studied, (2) their {\it appearance} along 
preferred directions in
the parameter space, serving as a guide to search for larger \RC s and
(3) for smaller dissipations the \RC\, increases significantly inside
the ISSs, even though they are not responsible for the larger \RC\,
observed, but the accelerator modes from the conservative limit.
Besides that it was also
observed that in chaotic regions the \RC s are smaller compared to the
regions with ISSs. The magnitude of the positive LEs also do not affect
the \RC. This complete connection between \RC s with a family of ISSs
and chaotic domains in parameter spaces gives us general clues for the
origin of directed
transport and further evidences of the generic properties of the ISSs
to generate such transport.

Besides the ratchet experiments with cold atoms 
\cite{tabosa99} or with the net motion of the particles in a 
silicon membrane with asymmetric pores \cite{matthias03}, 
suggested in \cite{alan11-1} as candidates to observe the generic 
ISSs experimentally, we would like to mention here another
recent experimental device for this purpose. It is the Leidenfrost
motion of solids and droplets \cite{lagubeau11,Stout} on 
a hot ratchet-like plate. We guess that determining the terminal 
velocity of the drops in the parameter space temperature of the 
plate against the radius of the drop, it should be possible to
see reminiscences of the ISSs.

\vspace*{-0.5cm}

%
\end{document}